\documentclass{article}
\usepackage{spconf,amsmath,graphicx}

\usepackage{enumitem}
\setlist{nosep, leftmargin=14pt}

\usepackage{xcolor}
\usepackage{url}
\usepackage{lipsum}
\usepackage{amsmath,amssymb,amsfonts}

\newcommand{\TODO}[1]{#1}
\DeclareMathOperator*{\argmin}{arg\,min}

\title{Deformable image registration with stochastically \\regularized biomechanical equilibrium}
\name{Pablo Alvarez $^{a, b, \star}$ \thanks{$\star$ Corresponding author: pablo.alvarez@inria.fr} and St\'ephane Cotin $^{a, b}$}
\address{$^{a}$ Inria, Strasbourg, France \\
    $^{b}$ ICube, University of Strasbourg, CNRS, Strasbourg, France}
\begin{document}
%
\maketitle
\begin{abstract}
Numerous regularization methods for deformable image registration aim at enforcing smooth transformations, but are difficult to tune-in \textit{a priori} and lack a clear physical basis. Physically inspired strategies have emerged, offering a sound theoretical basis, but still necessitate complex discretization and resolution schemes. This study introduces a regularization strategy that does not require discretization, making it compatible with current registration frameworks, while retaining the benefits of physically motivated regularization for medical image registration. The proposed method performs favorably in both synthetic and real datasets, exhibiting an accuracy comparable to current state-of-the-art methods.
\end{abstract}
\section{Introduction}
Deformable image registration (DIR) of medical images has been a subject of research for decades, given its great potential for clinical applications such as disease diagnosis \cite{Khalil2018} and interventional guidance \cite{Alvarez2021}, to name a few. Despite the important progress made in recent years, DIR methods still lack efficiency and robustness, which limits their transferability into clinical practice. This stems from the fact that DIR is an ill-posed problem, partly because of the characteristics of medical images (\textit{e.g.} low contrast and noise), but also because the underlying organs may deform considerably. To address this issue, multiple regularization strategies have been proposed \cite{Sotiras2013}, many of which were designed specifically to strengthen transformation smoothness. Although this is a desirable property, overly-smooth transformations can reduce registration accuracy, and properly adjusting regularization strength may become an important issue. Furthermore, it is difficult to provide a physical justification for some of these commonly used regularization strategies \cite{Burger2013}.

In response, physically motivated regularization has been introduced. These methods consider the organs in medical images as hyperelastic bodies, and exploit the theory of continuum mechanics for the design of regularization strategies \cite{Burger2013,Genet2018}. In addition to their strong physical basis, they also inherently strengthen desirable properties such as transformation smoothness and invertibility. However, to date, they still require complex domain discretization and resolution schemes. In this work, we extend on these ideas to propose a physically-derived regularization strategy that can be computed locally (pointwise) and does not require any discretization. As such, it facilitates integration with existing DIR frameworks and optimizers, while maintaining the desirable properties of physically motivated regularization.

\subsection{The ill-posed optimization problem}
Let $\mathcal{F}, \mathcal{M}$ 
be the \textit{d}-dimensional fixed and moving images, defined in domains $\Omega_0$ and $\Omega$, respectively. The registration problem consists in finding a displacement field $\mathbf{u}$, or equivalently, a transformation $\boldsymbol{\phi} : \mathbf{x} \in \Omega_0 \mapsto \mathbf{x} + \mathbf{u}(\mathbf{x})$, 
that aligns structures in $\Omega_0$ to homologous structures in $\Omega$, \textit{i.e.} $\mathcal{F}(\mathbf{x})$ and $\mathcal{M}(\boldsymbol{\phi}(\mathbf{x}))$ are close under some similarity criterion. This is typically formulated as an optimization problem:
\begin{equation}
    \boldsymbol{\hat{\phi}} = \argmin_{\boldsymbol{\phi}} \, (1 - \beta) \, \mathcal{L}_{sim}(\boldsymbol{\phi}; \mathcal{F}, \mathcal{M}) + \beta \, \mathcal{L}_{reg}(\boldsymbol{\phi})
    \label{eq:optim_problem}
\end{equation}
where $\mathcal{L}_{sim}$ is a metric measuring the similarity between the images, $\mathcal{L}_{reg}$ is a regularization term and $\beta$ is a normalized weighting coefficient. 

In general, DIR is ill-posed, and many similarity and regularization terms have been proposed in an effort to overcome this problem \cite{Sotiras2013}. In this work, we focus our attention on the regularization term $\mathcal{L}_{reg}$, which is of great importance for high quality and physically plausible registration results.

\subsection{Regularization}
In medical image registration, it is important that the estimated transformation $\boldsymbol{\phi}$ is diffeomorphic, since it is the minimum requirement for physical plausibility. In practice, this implies that $\boldsymbol{\phi}$ is both smooth and invertible. Various regularization strategies have been proposed to directly reinforce these properties. For instance, the diffusion \cite{Fischer2002}, curvature \cite{Fischer2003} and bending energy \cite{Rueckert1999} regularizers were designed to strengthen smoothness. However, in the context of deforming organs, their rather topological motivation is difficult to justify in physical terms. 
An alternative approach is to consider the underlying deforming objects as elastic bodies, and directly use their strain energy to regularize the transformation $\boldsymbol{\phi}$, on linearly-elastic or hyperelastic settings \cite{Burger2013}. However, using strain energy as regularization penalizes deformation, since only rigid body motion yields zero strain energy. A more recent idea is to consider DIR as an elastostatic problem, and penalize any deviation from the solution of a hyperelastic body in equilibrium with arbitrary boundary tractions, but without body forces \cite{Genet2018}. This equilibrium gap regularization principle, allows for large deformations, provided that they are compatible with the conservation of linear momentum of the above-mentioned elastostatic problem:
\begin{equation}
  \nabla \cdot \left(\frac{\partial \Psi}{\partial \mathbf{F}}\right) = \nabla \cdot \mathbf{P} = \mathbf{0}, 
  \label{eq:equilibrium}
\end{equation}
with $\Psi$ the strain energy density function, $\mathbf{F} = \nabla \boldsymbol{\phi}$ the deformation gradient tensor and $\mathbf{P}$ the first Piola-Kirchhoff stress tensor. 
As such, regularization with the equilibrium gap principle is achieved by defining
\begin{equation}
    \mathcal{L}_{reg} = \left\lVert \nabla \cdot \mathbf{P} \right\rVert^2.
    \label{eq:equilibrium_gap}
\end{equation}
To the best of our knowledge, only Finite Element (FE) formulations of this regularization approach exist to date \cite{Genet2018,Lee2019}. As with the FE method in general, the accuracy depends on the quality and level of discretization, and \eqref{eq:equilibrium_gap} can only be enforced weakly. In this work, we propose a local (pointwise) implementation of the equilibrium gap regularization in \eqref{eq:equilibrium_gap} that does not require any FE discretization and can thus be enforced strongly. 

\section{Materials and Methods}
\subsection{Analytically regularizing the equilibrium gap}
Let $\Psi$ be the strain energy density function of an isotropic, homogeneous, compressible Neohookean material:
\begin{equation}
    \Psi = \frac{\lambda}{4}(J^2 - 1 - 2\,\mathrm{ln}(J)) + \frac{\mu}{2}(I_C - 3 - 2\,\mathrm{ln}(J)),
    \label{eq:strain_energy}
\end{equation}
with $I_C = \mathrm{tr}(\mathbf{C})$ the first invariant of the right Cauchy-Green deformation tensor $\mathbf{C} = \mathbf{F}^T \, \mathbf{F}$, $J = \mathrm{det}(\mathbf{F})$, and $\mu$ and $\lambda$ the so-called \textit{Lam\'e parameters}.

From the constitutive equation \eqref{eq:strain_energy}, we first derive an expression for the stress of the material,
\begin{equation}
    \mathbf{P} = \frac{\partial \Psi}{\partial \mathbf{F}} = \frac{\lambda}{2} (J^2 - 1) \mathbf{F}^{-T} + \mu (\mathbf{F} - \mathbf{F}^{-T}),
\end{equation}
and then take its divergence to rewrite \eqref{eq:equilibrium} as: 
\begin{eqnarray}
    \nabla \cdot \mathbf{P} &=& \frac{\lambda}{2} \left(2J\mathbf{F}^{-T}\nabla J + (J^2 - 1)\nabla \cdot \mathbf{F}^{-T}\right) \nonumber \\
    &&{+}\: \mu\left(\nabla \cdot \mathbf{F} - \nabla \cdot \mathbf{F}^{-T}\right) = \mathbf{0}. \label{eq:equilibrium_analytical}
\end{eqnarray}
It should be noted that the computation of \eqref{eq:equilibrium_analytical} at a material point necessitates the first- and second-order derivatives of the displacement field $\mathbf{u}$. In a FE formulation, this is possible through the so-called shape functions of the FE spaces. Here, we suppose that analytical expressions for such derivatives are available, and therefore the computation of the equilibrium gap \eqref{eq:equilibrium_gap} can be performed directly. This is the case, for instance, for deformable image registration methods based on parametric transformations \cite{Rueckert2006}. In this work, we exploit the recently developed Implicit Neural Representations for deformable image registration (IDIR) for our implementation. 

\subsection{Images and validation data}
\subsubsection{Synthetic hyperelasticity dataset}
This dataset consists of three 2-dimensional deformation scenarios of a hyperelastic unit square constrained at its boundary with randomly generated, smooth displacements. The variational formulation of the elastostatic problem (under plane strain assumptions) described by \eqref{eq:equilibrium} and \eqref{eq:strain_energy} was implemented in FeniCs \cite{Logg2012}. 
The displacement fields for the Dirichlet boundary conditions were generated from the sum of 12 randomly parameterized Gaussian radial basis functions with experimentally defined parameter bounds. 
The simulations were performed with a FE $40 \times 40$ grid mesh of quadratic quad elements, containing a total of 1600 elements and 14641 mesh nodes. We considered a highly compressible material with a unit Young's modulus and a Poisson ratio of 0.3, yielding approximately to $\mu=0.577$ and $\lambda=0.385$.
\begin{figure}[tb]
    \centerline{\includegraphics[width=0.85\columnwidth]{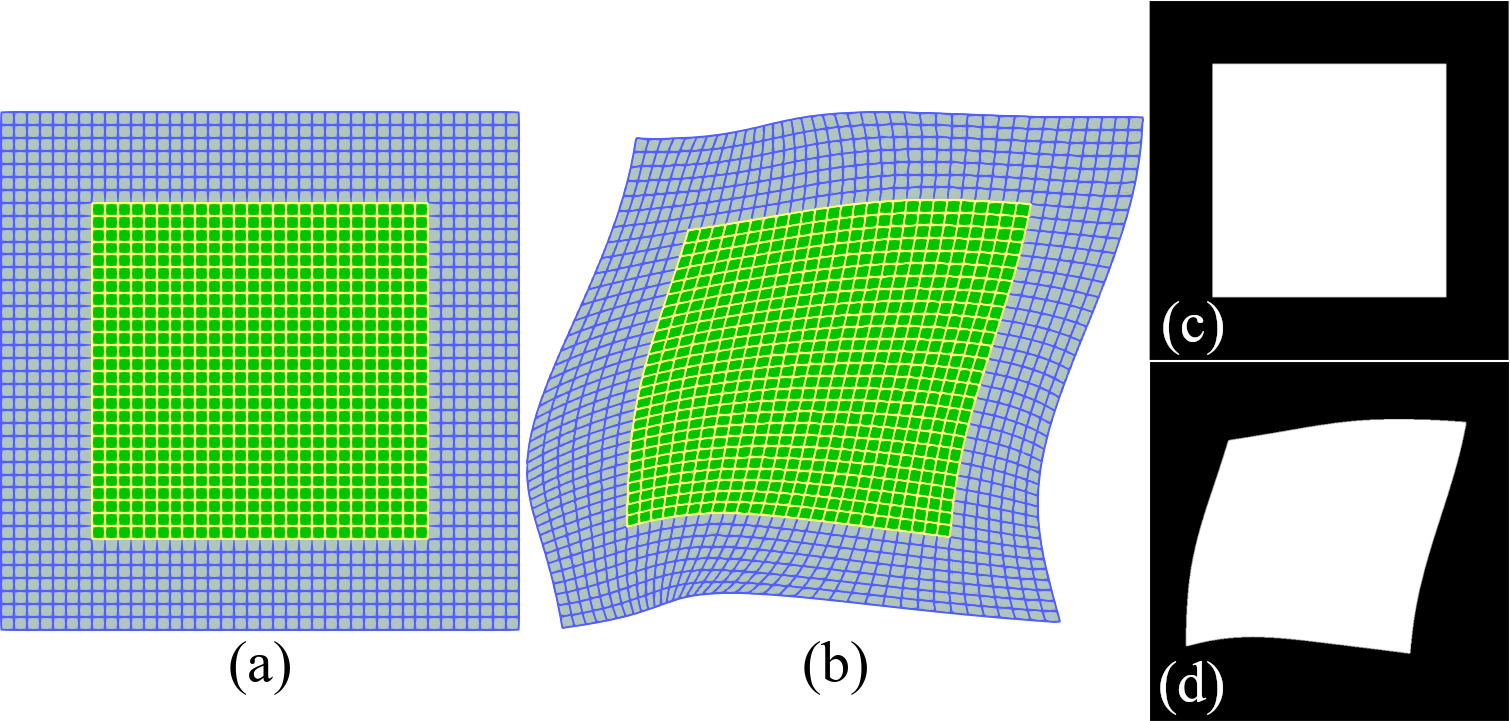}}
    \caption{Synthetic dataset generation. FE meshes in undeformed (a) and deformed (b) configurations for a hyperelastic unit square with random Dirichlet boundary conditions. The generated fixed (c) and moving (d) binary images. \label{fig:synthetic_dataset}}
\end{figure}
For each deformation scenario, fixed and moving binary images were generated from the solution of the elastostatic problem as illustrated in Fig. \ref{fig:synthetic_dataset}. 
Note that similar 2D binary datasets have been used in the past for validating image registration methods, \textit{e.g.} \cite{Burger2013}. Our dataset, however, differs from these works in two key aspects: (i) the object is considered hyperelastic, which entails some constraints in the resulting deformation fields; and (ii) the ground truth displacement is known at each mesh node, which facilitates quantitative evaluation of registration performance. 
\begin{figure}[tb]
    \centerline{\includegraphics[width=0.95\columnwidth]{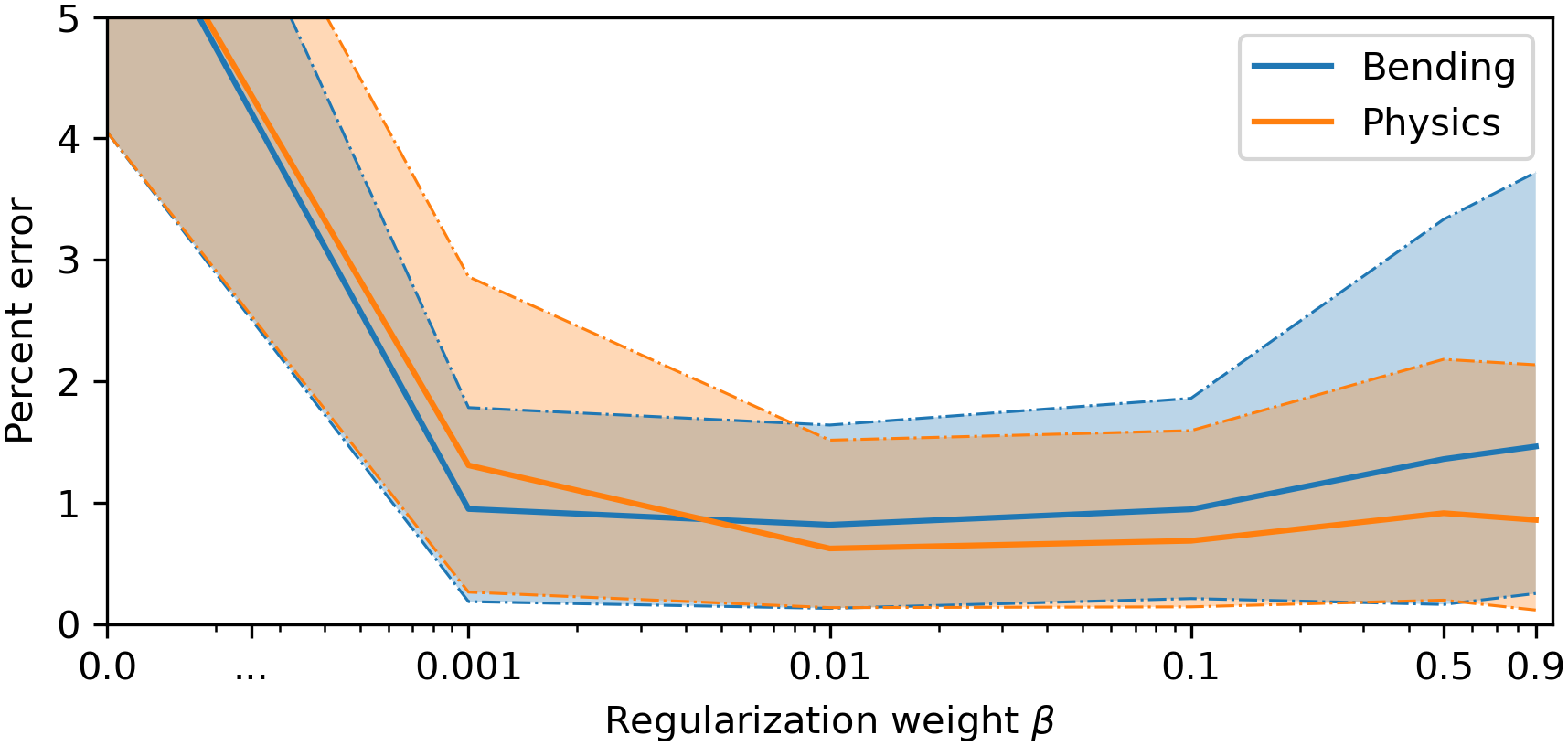}}
    \caption{Registration accuracy on synthetic dataset for bending energy regularization (Bending) and the proposed \TODO{analytical equilibrium gap} regularization (Physics) with varying $\beta$, across deformation scenarios. Mean (solid line), first quantile (lower bound) and third quantile (upper bound) of registration error as percentage of unit length. \label{fig:regweight_synthetic}}
\end{figure}


\subsubsection{DIRLab 4DCT dataset}
The DIRLab 4DCT dataset \cite{Castillo2009} contains ten cases of lung CT images taken at various phases of a respiratory cycle, with varying image resolutions ranging from $256 \times 256$ to $512 \times 512$ for the in-plane resolution and 256 or 512 for the number of slices. As in previous works, we evaluated registration performance only using the end-inhalation (fixed) and end-exhalation (moving) images, for which a set of 300 paired anatomical landmarks is available. In addition, we extracted lung masks from each image using an automatic method \cite{Hofmanninger2020} in order to define regions of interest for transformation optimization, as is common practice in the literature \cite{Ruhaak2017,Wolterink2022}. 

\subsection{Implementation within the IDIR framework}
Recently, implicit neural representations for deformable image registration (IDIR) have been proposed \cite{Wolterink2022,Harten2023}. These learning techniques implicitly represent the transformation $\boldsymbol{\phi}$ through neural networks, using as input spatial coordinates $\mathbf{x}$, and as output the corresponding displacements $\textbf{u}$. 
When combined with periodic activation functions, these networks allow the computation of high-order spatial derivatives of the displacement field $\mathbf{u}$, 
therefore providing a framework for the implementation of complex regularization schemes as the proposed \TODO{analytical equilibrium gap regularizer}.

We implemented the IDIR framework in PyTorch following \cite{Wolterink2022}. Given the invariance of IDIR to image resolution, we used the same architecture for all registration problems: 3 hidden layers of 256 sine-activated units and \textit{d}-dimensional input/output layers, accordingly. 
The network weights were initialized following \cite{Sitzmann2020}. 
For each registration task, the optimization problem \eqref{eq:optim_problem} was solved stochastically using the Adam optimizer and 10000 randomly sampled points, 
pooled from the whole fixed domain for the synthetic dataset, and from the lung masks for the DIRLab 4DCT dataset. 

The mean squared error and normalized cross-correlation were used as $\mathcal{L}_{sim}$ for the synthetic and DIRLab 4DCT datasets, respectively. For the regularization $\mathcal{L}_{reg}$, the bending energy \cite{Rueckert1999} and our proposed \TODO{analytical equilibrium gap regularizer} were implemented. For the latter, we chose a unit Young's modulus and zero Poisson's ratio, yielding $\mu=\frac{1}{2},\lambda=0$. This choice reduces the registration bias for the synthetic dataset (same hyperelastic model but different parameters), and also avoids over-penalization of volume changes that are expected in the DIRLab 4DCT dataset. 


\section{Results}
\subsection{Synthetic hyperelasticity dataset}
\begin{figure}[tb]
    \centerline{\includegraphics[width=0.95\columnwidth]{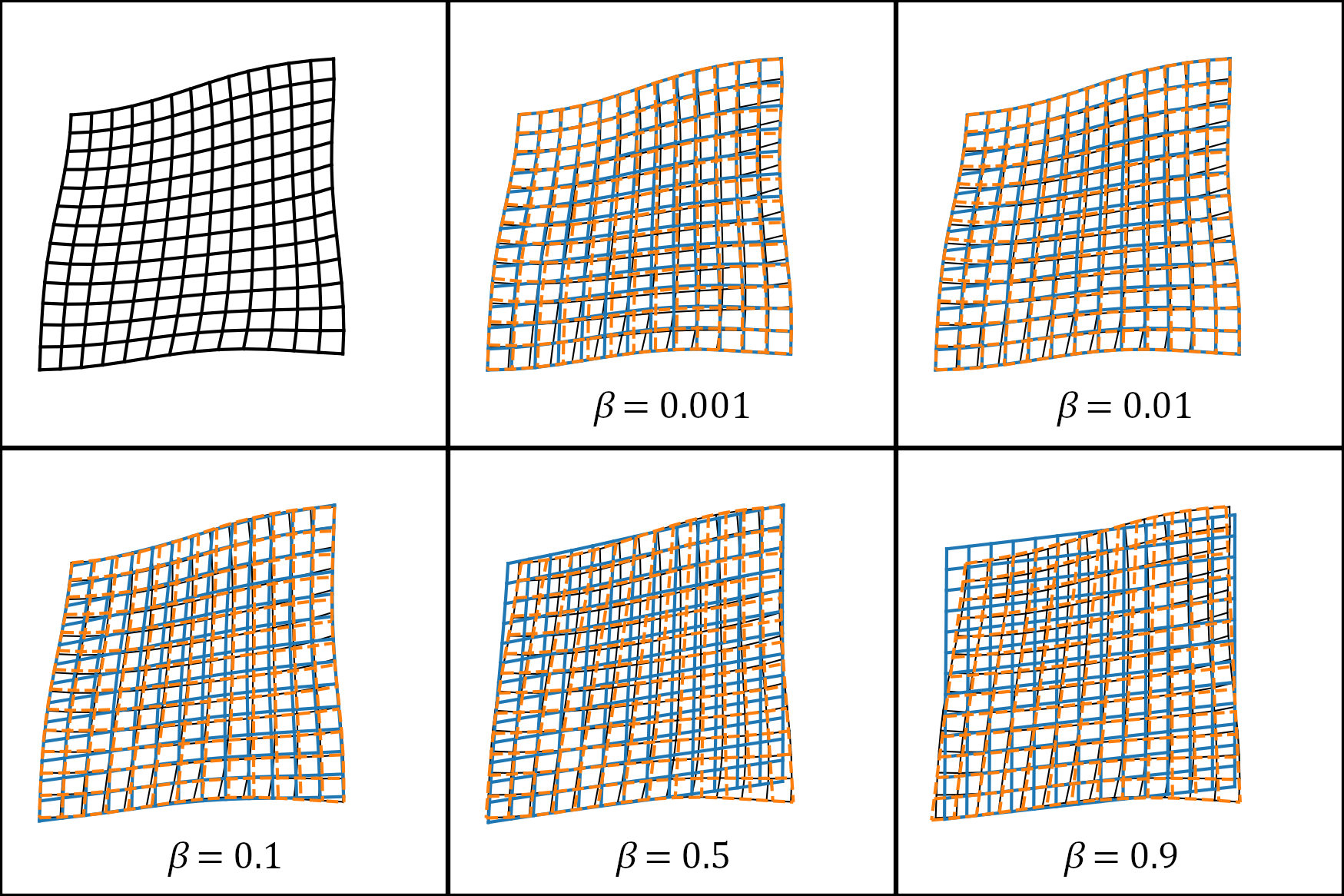}}
    \caption{Qualitative registration results on synthetic dataset for bending energy regularization (solid blue) and the proposed \TODO{analytical equilibrium gap} regularization (dashed orange) with varying $\beta$, on top of the ground truth deformed geometry (solid black). 
    \label{fig:regweight_synthetic_qualitative}}
\end{figure}

Considering that the functional of the optimization problem \eqref{eq:optim_problem} is a convex combination of the similarity $\mathcal{L}_{sim}$ and regularization $\mathcal{L}_{reg}$ terms, the regularization weight $\beta$ affects both. Therefore, we chose values spanning the unit interval to evaluate the influence of the regularization weight in registration accuracy, namely $\beta \in \{0.0, 0.001, 0.01, 0.1, 0.5, 0.9\}$. In Fig. \ref{fig:regweight_synthetic}, the distribution of percent error across all deformation scenarios is presented; and, to further illustrate such influence, Fig. \ref{fig:regweight_synthetic_qualitative} depicts the result of forward-warping a uniform grid with the estimated transformations for one of the three deformation scenarios.

\subsection{DIRLab 4DCT dataset}
We evaluated registration accuracy in the DIRLab dataset using snap-to-voxel target registration errors (TRE) for the 300 landmarks available per case, for both bending energy and our proposed \TODO{ analytical equilibrium gap regularization} strategies. The results of this quantitative evaluation are presented in Table \ref{tab:tre_dirlab}, along with those for the current state-of-the-art method in the DIRLab dataset \cite{Vishnevskiy2017} for comparison. The regularization weight for this evaluation was set to $\beta=0.001$, which was found to provide optimal results across cases. 
\begin{table}[htb]
    \caption{Mean (standard deviation) snap-to-voxel TRE (mm) with bending energy (Bending) and the proposed \TODO{analytical equilibrium gap} (Physics) regularizers, along with the state-of-the-art method for the 4DCT DIRLab dataset. \label{tab:tre_dirlab}}
    \begin{center}
    \begin{tabular}{cccc}
    \textbf{Scan} & \textbf{Bending} & \TODO{\textbf{Physics}} & \textbf{isoPTV} \cite{Vishnevskiy2017} \\
    \hline
    4DCT 01 & 0.77 (0.90) & 0.79 (0.93) & 0.76 (0.90) \\
    4DCT 02 & 0.69 (0.88) & 0.78 (0.91) & 0.76 (0.89) \\
    4DCT 03 & 0.91 (1.03) & 0.90 (1.05) & 0.90 (1.05) \\
    4DCT 04 & 1.35 (1.22) & 1.33 (1.27) & 1.24 (1.29) \\
    4DCT 05 & 1.27 (1.49) & 1.22 (1.50) & 1.12 (1.44) \\
    4DCT 06 & 1.12 (1.06) & 1.13 (1.05) & 0.85 (0.89) \\
    4DCT 07 & 1.10 (0.98) & 1.05 (0.98) & 0.80 (1.28) \\
    4DCT 08 & 1.29 (1.36) & 1.24 (1.28) & 1.34 (1.93) \\
    4DCT 09 & 1.17 (0.99) & 1.17 (1.03) & 0.92 (0.94) \\
    4DCT 10 & 1.13 (1.05) & 1.12 (1.06) & 0.82 (0.89) \\
    \hline
    Average & 1.19 & 1.18 & 0.95
    \end{tabular}
    \label{tab1}
    \end{center}
\end{table}

\begin{figure}[tb]
    \centerline{\includegraphics[width=0.95\columnwidth]{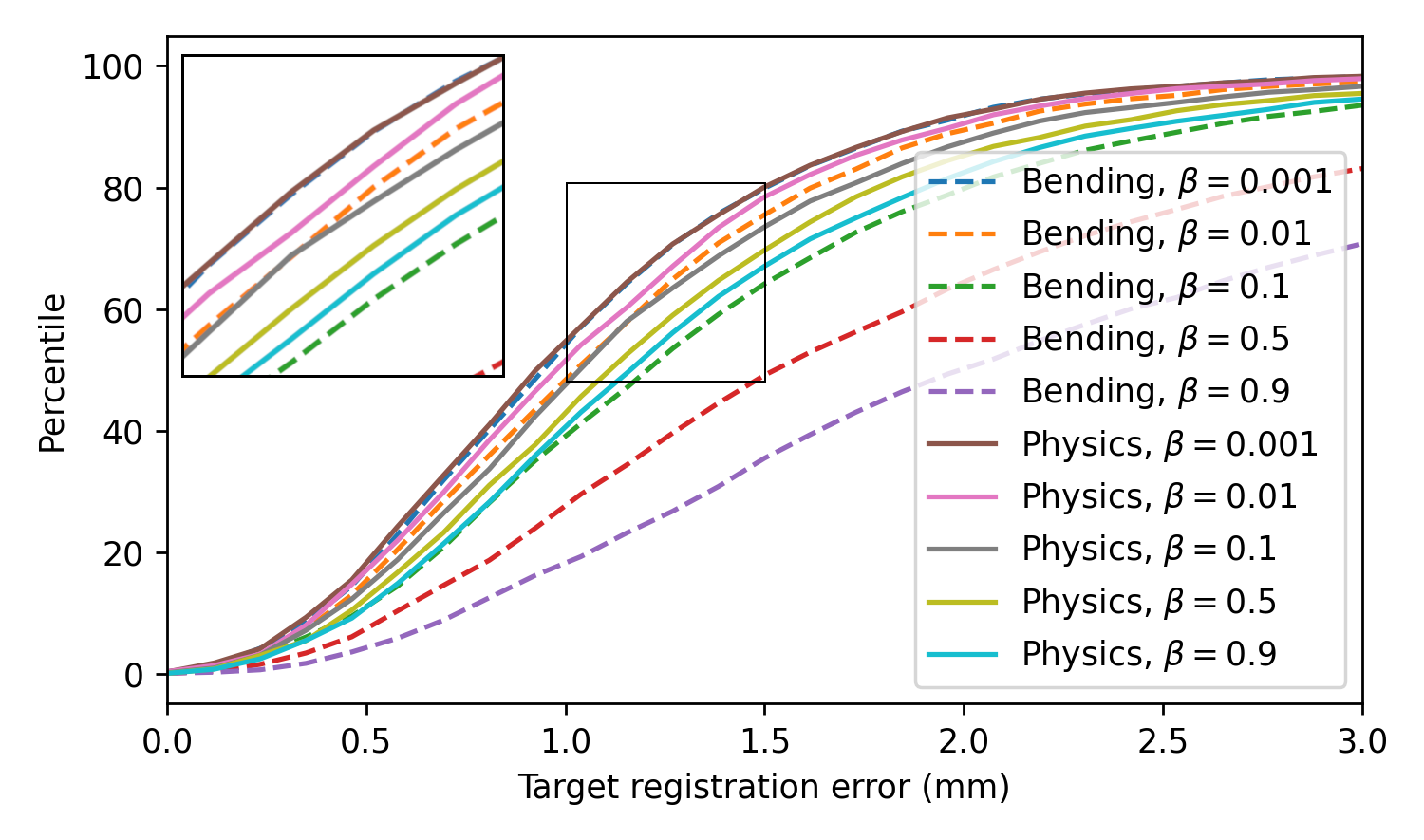}}
    \caption{Cumulative distribution of TRE with bending energy regularization (Bending) and the proposed \TODO{analytical equilibrium gap} regularization (Physics) with varying $\beta$ for the DIRLab 4DCT dataset. \label{fig:tre_cum_dirlab}}
\end{figure}

The effect of regularization weight $\beta$ was also evaluated for the DIRLab dataset, but in terms of cumulative TRE, as in previous works \cite{Ruhaak2017,Wolterink2022}. These results are depicted in Fig. \ref{fig:tre_cum_dirlab}. 

\section{Discussion and conclusion}
The proposed \TODO{analytical equilibrium gap} regularization strategy resulted in comparable or even improved registration accuracy than bending energy regularization, for our experiments with synthetic and DIRLab 4DCT datasets. Both regularization strategies performed well in the evaluated datasets, but the proposed regularization was more robust to changes in regularization weight $\beta$. 
This is clearly illustrated in Figs. \ref{fig:regweight_synthetic} and \ref{fig:tre_cum_dirlab}, where the registration accuracy was still acceptable with the proposed regularization strategy even for very high values of regularization weight ($\beta \in \{0.5, 0.9\}$), while being comparable with the state-of-the art in the DIRLab 4DCT dataset for an optimal weight ($\beta = 0.001$, see Table \ref{tab:tre_dirlab}). In practice, it is impossible to choose \textit{a priori} a value for $\beta$ yielding optimal accuracy, and therefore, the lesser its influence in registration accuracy, the better. We want to emphasize that the bending energy regularization, as well as other non-physically motivated regularization strategies, may, by construction, over-regularize the transformation $\boldsymbol{\phi}$ in a large deformation setting. This can be seen in Fig. \ref{fig:regweight_synthetic_qualitative}, where a decrease in deformation with increasing $\beta$ is observed for bending energy regularization, effect that is significantly lower for our proposed \TODO{physically motivated regularization}. In fact, with high regularization weight, the estimated transformation approaches an affine transformation, which is not penalized by this type of regularization. 

There are, however, some limitations that should be addressed in future work. A first limitation concerns our current implementation, which depends on the IDIR framework. Although we tried to limit the impact of the network hyperparameters by reutilizing those proposed in a previous study \cite{Wolterink2022}, these may affect registration accuracy and a sensibility study is therefore welcomed. We recall, however, that the IDIR framework was used herein only as a proxy for studying the performance of our \TODO{proposed regularization strategy}, and other registration approaches could be used, provided that the first and second derivatives of the displacement field are available. For instance, we plan an implementation in the parametric image registration framework Elastix in the future. A second limitation is the comparison to bending energy regularization only. Nonetheless, as mentioned above, other non-physically inspired regularization strategies will most likely fail in large deformation settings (with high regularization weight) as they inherently penalize deformation. A more informative comparison would be to the equilibrated warping approach \cite{Genet2018,Lee2019}, since it uses the same mechanical quantity for regularization as in this work. The difference is that we propose a local (pointwise) form in \eqref{eq:equilibrium_analytical} optimized stochastically, as opposed to a variational form weakly optimized globally in \cite{Genet2018}. Although we do not require any discretization, it is unclear whether both strategies would yield the same solution, with the same efficiency and precision. We plan to investigate these considerations in future work.

Finally, extensions to the proposed regularization strategy are possible. For instance, other hyperelastic constitutive laws such as the Money-Rivlin or Yeoh models may be implemented similarly. Also, the tissue parameters in \eqref{eq:equilibrium_analytical} can be tuned in for specific applications, to enforce quasi-incompressibility, for example. We are confident that physically motivated regularization can lead to more meaningful registration methods for the solution of a variety of problems.

\section{Compliance with ethical standards}
This research study uses human data from the 10-years old DIRLab 4DCT dataset made available in open access \cite{Castillo2009}. The dataset contains retrospective data acquired with the approval of the local review board (RCR 03-0800). 

\bibliographystyle{IEEEbib}
\bibliography{biblio}

\end{document}